# A Deterministic Solution of the Wigner Transport Equation with Infinite Correlation Length

Kyoung Yeon Kim and Byung-Gook Park

Department of Electrical and Computer Engineering, Seoul National University, Seoul, South Korea

*Abstract*—We propose a new formulation of the Wigner transport equation with infinite correlation length. Since the maximum correlation length is not limited to a finite value, there is no uncertainty in the simulation results owing to the finite integral range of the nonlocal potential term. For general and efficient simulation, the WTE is solved self-consistently with the Poisson equation through the finite volume method and the fully coupled Newton-Raphson scheme. Through this, we implemented a quantum transport steady state and transient simulator with excellent convergence.

## I. Introduction

Semiclassical models based on Boltzmann transport equation (BTE) have been widely used for semiconductor device simulation [1]-[4]. However, as devices continue to scale down, it is important to consider the quantum mechanical effects such as tunneling and quantization [5]-[7]. Therefore, semiclassical models are no longer valid at gate lengths of 10 nm or less, and quantum transport models are required for accurate prediction of device characteristics prediction.

For quantum transport simulation, non-equilibrium Green's function method (NEGF), Wigner transport equation (WTE), and master equation approach are mainly used. Although these models successfully predict the characteristics of nanoscale devices, the long computational time and lack of versatility of the simulator are still a critical issue.

In NEGF formalism, it is hard to consider the microscopic scattering mechanism because it requires the inversion of matrix of huge rank because the self-energy terms are generally nonlocal function [11]. Electron-phonon scattering can be efficiently calculated through local approximation, but large computational cost is required to include other scattering mechanisms. Also, although this method is well defined in steady state, it is inconvenient for transient simulation yet which is very important in device characterization [12].

Recently, Pratik B. Vyas et al reported a simulation of the dissipative quantum transport through the Pauli master equation (PME) [13]. They show successful simulation results in an ultra-thin body double-gate FET based on the quantum transmitting boundary method (QTBM). This is an attractive model for efficiently handling the scattering mechanism, but it can be applied only when the perturbation is weak and the device length is sufficiently short [13].

As an alternative to the above two methods, we focus on the WTE for the simulation of quantum transport in this work [14]-[16]. Transient simulation and dissipative transport simulation are possible based on the WTE, and there have been several studies on this direction [17]-[22]. However, more proper formulation of WTE for practical simulations should be introduced. There is a nonlocal potential term in WTE, which includes an infinite range of integration (infinite correlation length):

$$\hat{V}(\chi,k) = \frac{1}{i}\int_{-\infty}^{\infty} d\zeta\, e^{i\zeta k}\left[U\left(\chi+\frac{1}{2}\zeta\right) - U\left(\chi-\frac{1}{2}\zeta\right)\right]. \qquad (1)$$

where $\chi$ is real space index, $k$ is momentum space index and $U$ is potential energy. To solve this numerically, previous studies have used finite correlation length, resulting in several problems. W. R. Frensley noted that if the computational domain is confined to a finite region, Wigner-Weyl transformation cannot be unitary, because some of the information will be lost. A. S. Costolandski et al confirmed that different simulation results were obtained depending on the correlation length, and mentioned that the appropriate correlation length is different depending on the device structure and there is no simple physics-based rule to determine it. As such, the simulation based on the finite correlation length has a pro blem in that there is uncertainty in the simulation result depending on the correlation length and may not be physically consistent with the density operator.

Recently, L. Schulz et al already proposed complex absorbing potential formalism to consider the unbounded computational domain. However, we present an alternative and much simpler method in this paper. We derive a new formulation with an infinite correlation length by assuming an ideal semi-infinite reservoir in the contact region. Through reconstruction of the nonlocal potential term, an equivalent equation with a finite integral range is derived. Since our new formula considers the integral range of nonlocal potential terms up to infinity, it can solve the problem of uncertainty of simulation results according to the integral range. To solve the proposed equation numerically, we use the unwind scheme, finite volume method (FVM), and backward Euler method. Through these, quantum transport steady-state and transient simulation with excellent convergence are successfully implemented. By applying our simulator to resonant tunneling diode (RTD), it was confirmed that reliable results are obtained by showing the plateau region and transient oscillation in unstable bias.

## II. Theory

The WTE can be expressed as follow:

$$\left(\frac{\partial f}{\partial t}\right)_C - \frac{\partial f}{\partial t} = \frac{\partial \varepsilon}{\hbar \partial k}\frac{\partial f}{\partial \chi} + \frac{1}{\hbar}\int_{-\infty}^{\infty}\frac{dk'}{2\pi}\hat{V}(\chi, k-k')f(\chi, k'), \quad (2)$$

where the first term of LHS is scattering integral, $t$ is time, $\hbar$ is Dirac's constant, and $\varepsilon$ is energy level. In general, to solve the WTE numerically, the integral range of the nonlocal potential term (Eq. (1)) is limited to a finite range.

To consider the infinite correlation length, we first assume an ideal contact condition in which the reservoir is semi-infinitely long and has a constant potential energy. Such boundary conditions are commonly used in quantum transport simulation. In a one-dimensional space, the contact is located at $\chi=0$ and $\chi=L$, and the potential energy at the right contact is higher than the left one by $U_{ex}$. To reconstruct the equation, the nonlocal potential term is divided into the sum of the two terms as follows:

$$\hat{V}(\chi,k) = \frac{1}{i}\int_{-\infty}^{\infty}d\zeta e^{i\zeta k}\left[U_{cor}(\chi,\zeta) - U_{ex}u(\zeta) + U_{ex}u(-\zeta)\right] + \frac{1}{i}\int_{-\infty}^{\infty}d\zeta e^{i\zeta k}\left[U_{ex}u(\zeta) - U_{ex}u(-\zeta)\right], \quad (3)$$

$$U_{cor}(x,\zeta) = U\left(\chi + \frac{1}{2}\zeta\right) - U\left(\chi - \frac{1}{2}\zeta\right), \quad (4)$$

where $u$ is a unit step function. In this representation, we just add and subtract the product of $V_{ex}$ and the unit step function to the nonlocal potential term. The integrand function of the first term in Eq. (3) becomes an odd function for $\zeta$, and the second term can be calculated through the Fourier transform relational expression of the unit step function as follow:

$$u(t) \xleftrightarrow{F.T.} \frac{1}{jw} + \pi\delta(w), \quad (5)$$

Thus, Eq. (3) can be rewritten as:

$$\hat{V}(\chi,k) = 2\int_0^{\infty}d\zeta \sin(\zeta k)\left[U_{cor}(\chi,\zeta) - U_{ex}u(\zeta)\right] + 2\frac{U_{ex}}{k}. \quad (6)$$

When $\chi$ is between 0 and L, the integrand of the first term is always 0 if $\zeta$ is greater than 2L. Therefore, the integration range can be reduced to [0, 2L]:

$$\hat{V}(\chi,k) = 2\int_0^{2L}d\zeta \sin(\zeta k)\left[U_{cor}(\chi,\zeta) - U_{ex}\right] + 2\frac{U_{ex}}{k} \quad (7)$$

Since the integral range is finitely limited through reformulation of nonlocal potential terms, it is possible to solve WTE with infinite correlation length numerically.

## III. Simulation methods

To solve our new formulation numerically, we use the finite volume method (box integration method). In the steady state, WTE can be expressed as

$$\frac{\partial \varepsilon}{\hbar \partial k}\frac{\partial}{\partial \chi}f(\chi,k)+W_{\chi,k}-C_{\chi,k}=0, \quad (8)$$

where W is quantum evolution term and C is collisional term. To apply the upwind scheme, the formula can be divided into two cases according to the direction of the group velocity:

$$\pm v_{\chi,k}\frac{\partial}{\partial \chi}f^{\pm}(\chi,k)+W^{\pm}_{\chi,k}-C^{\pm}_{\chi,k}=0, \quad (9)$$

where v is group velocity and the + sign represents when the group velocity is positive and negative, respectively. For easy box integration, the equation is transformed as follows using partial differentiation:

$$\pm \frac{\partial}{\partial \chi}v_{\chi,k}f^{\pm}(\chi,k) \mp f^{\pm}(\chi,k)\frac{\partial v_{\chi,k}}{\partial \chi}+W^{\pm}_{\chi,k}-C^{\pm}_{\chi,k}=0. \quad (10)$$

In x space with a uniform mesh size, the box integration at node xi can be obtained by integrating Eq. (13) from xi-1/2 to xi+1/2:

$$\pm \left[v_{\chi,k}f^{\pm}(\chi,k)\right]_{x_{i-0.5}}^{x_{i+0.5}} \mp \int_{x_{i-0.5}}^{x_{i+0.5}}\partial\chi f^{\pm}(\chi,k)\frac{\partial v_{\chi,k}}{\partial \chi}+(W^{\pm}_{\chi,k}-C^{\pm}_{\chi,k})\Delta x=0. \quad (11)$$

To calculate the first term of Eq. (14), we need to know the distribution function at xi and xj. The simplest way to do this is to use the average value of two adjacent nodes. However, in this work we use the Quadratic upstream interpolation for convective kinematics (QUICK) scheme for high numerical accuracy. The value at the cell face can be calculated as follows through the QUICK scheme:

$$\left[vf^{+}\right]_{x_{i-0.5}}=\frac{3}{4}\left[vf^{+}\right]_{x_{i-1}}+\frac{3}{8}\left[vf^{+}\right]_{x_{i}}-\frac{1}{8}\left[vf^{+}\right]_{x_{i-2}}, \quad (12a)$$

$$\left[vf^{+}\right]_{x_{i+0.5}}=\frac{3}{4}\left[vf^{+}\right]_{x_{i}}+\frac{3}{8}\left[vf^{+}\right]_{x_{i+1}}-\frac{1}{8}\left[vf^{+}\right]_{x_{i-1}}, \quad (12b)$$

$$\left[vf^{-}\right]_{x_{i-0.5}}=\frac{3}{4}\left[vf^{-}\right]_{x_{i}}+\frac{3}{8}\left[vf^{-}\right]_{x_{i-1}}-\frac{1}{8}\left[vf^{-}\right]_{x_{i+1}}, \quad (12c)$$

$$\left[vf^{-}\right]_{x_{i+0.5}}=\frac{3}{4}\left[vf^{-}\right]_{x_{i+1}}+\frac{3}{8}\left[vf^{-}\right]_{x_{i}}-\frac{1}{8}\left[vf^{-}\right]_{x_{i+2}}. \quad (12d)$$

If the calculation range is outside the boundary, it is assumed to have the same value as in the boundary. Dirichlet boundary conditions apply only to the left if the group rate is greater than 0, and only to the right if it is less than 0 as upwind method.

The second term of Eq. (14) vanishes if the group velocity is constant. However, if the group velocity changes at any point for reasons such as partial varying effective mass, the second term should be calculated. If the group velocity at xi abruptly changes from A to B, the second term can be written as

$$\mp \int_{x_{i-0.5}}^{x_{i+0.5}}\partial\chi f^{\pm}(\chi,k)(B-A)\delta(\chi-\chi_{i})=\mp(B-A)f^{\pm}(x_{i},k), \quad (13)$$

where a is the Dirac delta function which is the derivative of the unit step function. For example, assuming a parabolic band, if the effective mass at xi changes from m1 to m2, Eq. (17) becomes

$$\mp\left(\frac{\hbar k}{m_2}-\frac{\hbar k}{m_1}\right)f^{\pm}(x_i,k). \tag{14}$$

Before describing the quantum evolution term, mesh spacing in k-space should be considered. When Eq. (13) is integrated over k-space, the equation becomes a continuity equation for charge density. In order to satisfy the charge conservation, the integral of the quantum evolution term must also be 0. If a uniform mesh size is used, the integral can be expressed discretely as

$$\int W^{\pm}{}_{\chi,k}dk = \sum_i W^{\pm}{}_{\chi,k_i}\Delta k = 0. \tag{15}$$

The above equation holds when the Fourier completeness relation is satisfied, and the mesh size at that time can be expressed as follow:

$$\int W^{\pm}{}_{\chi,k}dk = \sum_i W^{\pm}{}_{\chi,k_i}\Delta k = 0. \tag{16}$$

The above equation holds when the Fourier completeness relation is satisfied, and the mesh size at that time can be expressed as follow:

$$\Delta k = \frac{\pi}{N_k \Delta x}, \tag{17}$$

where Nk is the number of meshes in k-space.

For calculating the quantum evolution term, we first need to calculate the nonlocal potential term in Eq (10). To calculate the first term of RHS, we assume that the potential changes linearly between meshes as shown in Fig. 2. In this way, the nonlocal potential is calculated through direct integration rather than discrete integration in order to accurately account for changes in the sin function by position. Since a linear potential is assumed, the equation can be expressed in the form of a product of a sin function and a linear function, so that analytical calculations are easily possible. Eventually, the quantum evolution term is calculated discretely as follow:

$$V(\chi,k) = \frac{\Delta k}{2\pi\hbar}\sum_i V(\chi,k-k_i)f(\chi,k_i) \tag{18}$$

To simply include the scattering effect we use the relaxation time approximation:

$$C^{\pm}{}_{\chi,k} \approx -\frac{1}{\tau}\left[f^{\pm}(\chi,k) - \frac{f_{eq}(\chi,k)}{\int dk' f_{eq}(\chi,k')}\int dk' f^{\pm}(\chi,k')\right] \tag{19}$$

Where t is the relaxation time and f is the local equilibrium distribution function. The equilibrium distribution function uses the solution when the applied bias is zero.

Electrostatic potential Vp is obtained through the Poisson equation as

$$\frac{d^2V_p}{dx^2} = -\frac{q}{\varepsilon}\left[N_d(x)-n(x)\right], \tag{20}$$

where Nd is the doping concentration and n(x) is the electron density obtained from the Wigner function:

$$n(x) = \frac{1}{2\pi}\int f(x,k)dk. \tag{21}$$

The potential energy v(x) used in the nonlocal potential term can be calculated as follow:

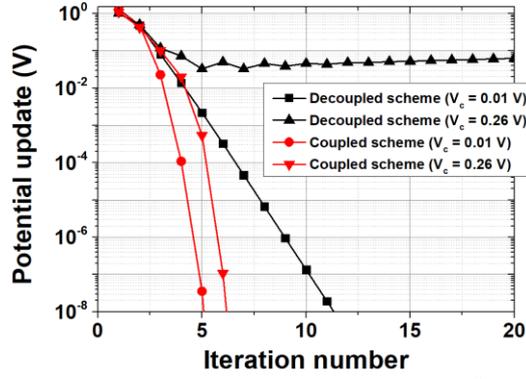

Fig. 1. Convergence of steady-state simulation according to iteration method. Coupled scheme (red line) shows better convergence than decoupled scheme (black line).

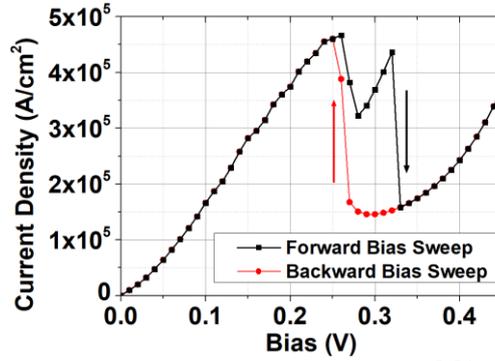

Fig. 2. Current-voltage characteristics obtained through steady-state simulation. 0.01V is used as the bias step, and bistability is shown in the case of forward bias sweep (black line) and backward bias sweep (red line).

$$v(x) = -qV_p(x) + U_c(x), \qquad (22)$$

where Uc is the band structure function which considers the band offset considering the barriers and wells.

Because an iterative solver is required because WTE is a nonlinear system that needs to be solved together with the Poisson equation, we simulate with two methods: the Gummel method [25], which is mainly used as a decoupled scheme, and the Newton Raphson method, which is a fully coupled scheme.

In the case of transient simulation, the calculation method is the same as that of steady-state simulation, and it is calculated using the backward Euler (implicit) method. Compared to the forward Euler (explicit) method, a much larger time step size can be used and more stable simulation is possible.

## IV. Simulation results

For the verification of our new model, we simulate a GaAs-AlGaAs-GaAs resonant tunneling diode as an example. The emitter and collector are 35 nm and have a doping concentration of $2*10^{18}$ /cm3, and the barriers and well are not doped. 3nm barriers, 4nm well, and 7 nm spacer are used. The band offset at the barrier is 0.3 eV and the temperature is 77 K. The relaxation time approximation is used to consider the scattering effects, and the relaxation time is 525 fs.

Fig. 1. shows the convergence of the simulation. At both low and high bias, the coupled method shows much better convergence. The simulation does not even converge with the decoupled method at 0.26 V. Of course, it is possible to enforce convergence by making the bias step smaller or multiplying the damping parameter, but the simulation efficiency deteriorates. Because the semiclassical assumption is used in the Gummel method, the convergence is poor in devices with strong quantum effect, but the Newton Raphson method shows quadratic convergence at all bias steps because it calculates an exact response matrix.

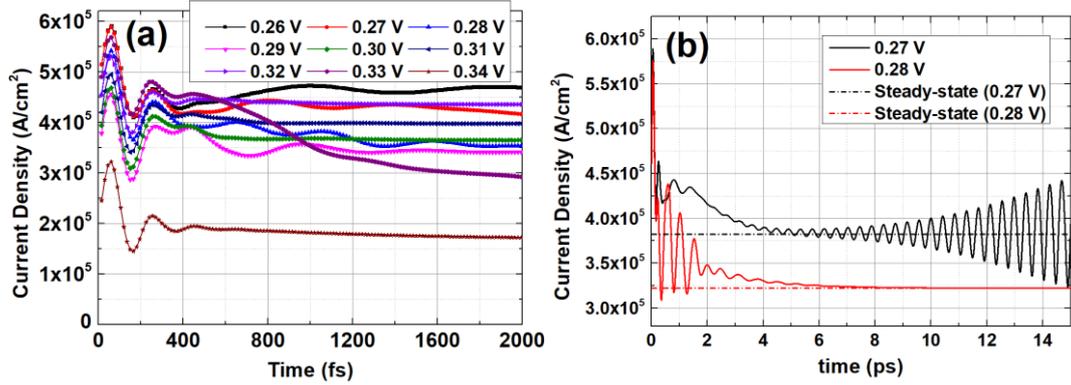

Fig. 3. Transient current characteristics up to (a) 2000 fs and (b) 15 ps. At all biases except 0.27 V, the current converges to a steady-state solution. (b) At 0.27 V, the current oscillates even after a long time, and the average value of the oscillation current is almost the same as that of the steady-state solution.

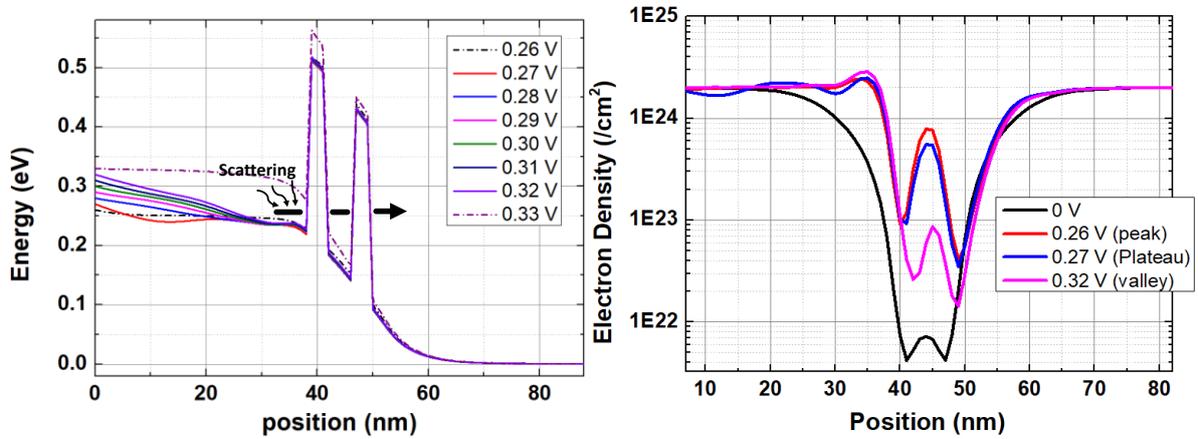

Fig. 4. (a) Conduction band diagram and (b) electron density in steady-state when forward bias sweep is performed.

The plateau region between 0.25 V and 0.32 V can be confirmed through the I-V characteristics in Fig. 2. A plateau appears in the case of a positive bias sweep. This result is obtained through the steady-state simulation method, so there is a possibility that an incorrect solution was found during the iteration process, so transient simulation is performed under the same conditions.

Fig. 3(a). shows the transient current characteristics when the bias is raised in the form of a step function by 0.01V by positive sweep. In all other biases except 0.27V, the current converges to the same value as the steady-state simulation result after a sufficient time as at 0.28 V in Fig. 3(b). However, the current oscillates at 0.27 V even after a long period of time. Specifically, it can be confirmed that only bias with negative resistance among the plateau region has unstable transient current characteristics.

Fig. 4 shows the band diagram at the plateau and the electron density at several biases. Fig. 4(a) shows the band diagram in steady-state when forward bias sweep is performed. The solid line from 0.27 V to 0.32 V is the plateau region, and we can know that the characteristic is clearly different from the normal operation region, which is the dotted line. At the plateau, we can see that band banding occurs in the emitter region in front of the first barrier. Therefore, the emitter region also shows characteristics like another quantum well, and a quantized state exists. And when this state is similar to the resonance energy level between the double barriers, a new current path is formed as shown schematically in Fig. 4(a). Therefore, as shown in Fig 4(b), not only the current in the plateau region but also the electron density in the quantum well does not drop significantly compared to the peak current.

Since our simulator clearly shows the unique characteristics of RTD such as plateau and this is consistent with previous studies [18]-[22], our new formulation is reliable and is expected to be used as a versatile quantum transport simulator.

## V. Conclusion

We derive a novel representation of nonlocal potential terms with infinite correlation length with the assumption of ideal contact. Through this, more accurate simulation is possible without uncertainty of the WTE solution due to the finite correlation length. Not only steady-state simulation but also transient simulation are possible, and since the Newton-Raphson method is used, the accurate linear response of the equation can be calculated, and thus small signal or noise analysis will also be readily possible. Now that the numerical properties and simulation results in one dimension have been confirmed, it is expected that it will be possible to extend to a three-dimensional simulation through a mode space approach, and it is expected that more realistic device simulation will be possible.

## Acknowledgement

This work was supported in part by the Brain Korea 21 FOUR Program of the Education and Research Program for Future ICT Pioneers, Seoul National University in 2020, in part by Future Semiconductor Device Technology Development Program (20010847 and 10067739) funded by Ministry of Trade, Industry and Energy (MOTIE) and Korea Semiconductor Research Consortium (KSRC), in part by National Research Foundation (NRF) funded by the Korean Ministry of Science and ICT under Grant 2020M3F3A2A01081670, 2020M3F3A2A01081666 and 2020R1A2C2103059, and in part by Synopsys INC.

## References

[1] S. Jin, T. Tang and M. V. Fischetti, "Simulation of Silicon Nanowire Transistors Using Boltzmann Transport Equation Under Relaxation Time Approximation," IEEE Transactions on Electron Devices, vol. 55, no. 3, pp. 727-736, Mar. 2008, DOI: 10.1109/TED.2007.913560.

[2] S. -M. Hong and C. Jungemann, "A fully coupled scheme for a Boltzmann-Poisson equation solver based on a spherical harmonics expansion," Journal of Computational Electronics 8, pp. 225-241, Oct. 2009, DOI: https://doi.org/10.1007/s10825-009-0294-y.

[3] S. Cha and S. -M. Hong, "Theoretical Study of Electron Transport Properties in GaN-Based HEMTs Using a Deterministic Multi-Subband Boltzmann Transport Equation Solver," IEEE Transactions on Electron Devices, vol. 66, no. 9, pp. 3740-3747, Sept. 2019, DOI: 10.1109/TED.2019.2926857.

[4] C. Jungemann, A. T. Pham, B. Meinerzhagen, C. Ringhofer, and M. Bollhofer, "Stable discretization of the Boltzmann equation based on spherical harmonics, box integration, and a maximum entropy dissipation principle," Journal of Applied Physics, Volume 100, Issue 2, July. 2006, DOI: https://doi.org/10.1063/1.2212207.

[5] Jing Wang and M. Lundstrom, "Does source-to-drain tunneling limit the ultimate scaling of MOSFETs?," Digest. International Electron Devices Meeting, 2002, pp. 707-710, DOI: 10.1109/IEDM.2002.1175936.

[6] D. Yadav and D. R. Nair, "Impact of Source to Drain Tunneling on the Ballistic Performance of Si, Ge, GaSb, and GeSn Nanowire p-MOSFETs," IEEE Journal of the Electron Devices Society, vol. 8, pp. 308-315, 2020, DOI: 10.1109/JEDS.2020.2980633.

[7] K. -H. Kao et al., "Subthreshold Swing Saturation of Nanoscale MOSFETs Due to Source-to-Drain Tunneling at Cryogenic Temperatures," IEEE Electron Device Letters, vol. 41, no. 9, pp. 1296-1299, Sept. 2020, DOI: 10.1109/LED.2020.3012033.

[8] F. Bonani, G. Ghione, M. R. Pinto and R. K. Smith, "An efficient approach to noise analysis through multidimensional physics-based models," IEEE Transactions on Electron Devices, vol. 45, no. 1, pp. 261-269, Jan. 1998, DOI: 10.1109/16.658840.

[9] C. H. Park and Y. J. Park, "Modeling of thermal noise in short-channel MOSFETs at saturation, Solid-state Electronics, vol. 44, issue 11, pp 2053-2057, Nov. 2000, DOI: https://doi.org/10.1016/S0038-1101(00)00161-1

[10] S. Hong and J. Jang, "Transient Simulation of Semiconductor Devices Using a Deterministic Boltzmann Equation Solver," IEEE Journal of the Electron Devices Society, vol. 6, pp. 156-163, 2018, DOI: 10.1109/JEDS.2017.2780837.

[11] P. Mahdi, "Numerical Study of Quantum Transport in Carbon Nanotube-based Transistors," dissertation, Institute for Microelectronics, Vienna University of Technology, Vienna, 2007.

[12] M. R. Hirsbrunner, T. M. Philip, B. Basa, Y. Kim, M. J. Park, and M. J. Gilbert, "A review of modeling interacting transient phenomena with non-equilibrium Green functions," Reports on Progress in Physics 82, 046001, March. 2019, DOI: https://doi.org/10.1088/1361-6633/aafe5f.

[13] P. B. Vyas, M. L. Van de Put, and M. V. Fischetti, "Master-Equation Study of Quantum Transport in Realistic Semiconductor Devices Including Electron-Phonon and Surface-Roughness Scattering," Physical Review Applied 13, 014067, Jan. 2020, DOI: https://doi.org/10.1103/PhysRevApplied.13.014067.

[14] E. Wigner, "On the Quantum Correction for Thermodynamic Equilibrium", Physical Review, 40, 749, June. 1932, DOI: https://doi.org/10.1103/PhysRev.40.749.

[15] M. Hillery, R. F. O'Connell, M. O. Scully, and E. P. Wigner, "Distribution Functions in Physics: Fundamentals", Physics Reports, vol 106, issue 3, pp 121-167, April. 1984, DOI: https://doi.org/10.1016/0370-1573(84)90160-1.

[16] W. R. Frensley, "Boundary Conditions for Open Quantum Systems Driven Far from Equilibrium," Reviews of Modern Physics. 63, 215, July. 1990, DOI: https://doi.org/10.1103/RevModPhys.62.745.

[17] S. Barraud, "Dissipative quantum transport in silicon nanowires based on Wigner transport equation," Journal of Applied Physics, vol 110, issue 9, Nov. 2011, DOI: https://doi.org/10.1063/1.3654143.

[18] K. L. Jensen and F. A. Buot, "Numerical simulation of intrinsic bistability and high-frequency current oscillations in resonant tunneling structures," Physical Review Letters. 66, 1078, Feb. 1991, DOI: https://doi.org/10.1103/PhysRevLett.66.1078.

[19] P. Zhao, H. L. Cui, and D. L. Woolard, Physical Review B. 63, 075302, Jan. 2001, DOI: https://doi.org/10.1103/PhysRevB.63.075302.

[20] B. A. Biegel, "Wigner Function Simulation of Intrinsic Oscillations, Hysteresis, and Bistability in Resonant Tunneling Structures," Porc. SPIE 3277, Ultrafast Phenomena in Semiconductors, April. 1998, DOI: https://doi.org/10.1117/12.306152.

[21] B. A. Biegel and J. D. Plummer, "Comparison of self-consistency iteration options for the Wigner function method of quantum device simulation," Physical Review B. 54, 8070, Sept. 1996, DOI: https://doi.org/10.1103/PhysRevB.54.8070.

[22] B. A. Biegel and J. D. Plummer, "Applied bias slewing in transient Wigner function simulation of resonant tunneling diodes," IEEE Transactions on Electron Devices, vol. 44, no. 5, pp. 733-737, May 1997, DOI: 10.1109/16.568033.

[23] B. P. Leonard, "A stable and accurate convective modelling procedure based on quadratic upstream interpolation," Computer Methods in Applied Mechanics and Engineering, vol 19, issue 1, pp 59-98, June 1979, DOI: https://doi.org/10.1016/0045-7825(79)90034-3.

[24] J. Butcher, Numerical Method for Ordinary Differential Equations, Wiley, 2003.

[25] H. K. Gummel, "A self-consistent iterative scheme for one-dimensional steady state transistor calculations," IEEE transactions on Electron Devices, vol 11, issue 10, pp 455-465, Oct 1964, DOI: 10.1109/T-ED.1964.15364.